# Making Design Practices Visible to Young Learners

**Rama Adithya Varanasi**
Information Science &
Technology
Pennsylvania State University
merama@protonmail.com

**Shulong Yan**
College of Education
Pennsylvania State University
suy114@psu.edu

**Dhvani Toprani**
College of Education
Pennsylvania State University
dhvanitoprani@psu.edu

**Marcela Borge**
College of Education
Pennsylvania State University
mborge@psu.edu

**ABSTRACT**

The role of design in K-12 education has increased in recent years. We argue that many of these design experiences do not help develop important habits of mind associated with Human Centered Design (HCD). In this paper, we present an approach for developing higher-order thinking processes associated with HCD as part of embedded design practice - an approach for teaching design thinking to younger children using principles of cognitive apprenticeship. First, we identify fundamental design habits of mind, discuss why it is difficult for young learners to develop such habits, and then draw upon cognitive apprenticeship principles to propose a concrete approach for design education. Finally, we present an illustration of embedded design practice to show how the situated context offers opportunities for designers to learn more about the needs of young learners while providing learners with opportunities to learn more about design practices.

**INTRODUCTION**

Design has become increasingly present in K-12 educational environments. The Next Generation Science standards call for the development of design thinking skills in K-12 environments [21]. As a result there has been an increasing emphasis on design thinking and pressure for teachers and informal learning environments to provide opportunities for students to engage in design. Given the growing emphasis on design thinking, it is not surprising that we have seen a recent surge in design thinking curriculum, technological tools, and makerspaces. However, the problem is that design, especially Human Centered Design, is a complex process that requires sophisticated ways of thinking which are often developed alongside technical skills. While there are many methods and approaches to teach older students about Human Centered Design, many questions remain with regard to how we develop these design skills in elementary students. In this paper, we present an approach to help enculturate students towards design ways of being that draws on theories of cognitive apprenticeship and Human-Centered design. We call this approach embedded design practice (EDP). We believe embedded design practice can provide quality learning outcomes for learners and designers.

In our previous work, we have written about how we get students to engage in complex forms of reasoning and negotiation in after-school design contexts [Authors citation]. We do so by empowering learners to be co-designers of their own design club [Authors citation]. By creating a narrative of the facilitators as designers of such experience and the students as clients and co-designers, we are able to apply cognitive apprenticeship methods of modeling, scaffolding, reflection, and exploration in a situated context.

The following work focuses on another way of embedding real-world design practices within educational contexts, while still adhering to principles of cognitive apprenticeship. Our aim is to develop models of design education that help students form essential design habits of mind: habits associated with complex forms of design thinking.

We begin this paper by identifying essential design habits of mind and discuss why such habits may be difficult for young learners to develop. We then discuss cognitive apprenticeship as a means to help develop habits of mind and build upon it to propose a concrete approach for design education. Finally, we present an embedded study conducted as part of a design learning activity to showcase how such embedded learning experiences can help us to better understand the needs of young learners while providing situated contexts for instruction. We conclude our paper by discussing implications for teaching and research.

i

## RELATED WORK

### Design habits of mind

Effective design is more than making and tinkering, it is a form of science inquiry that includes both divergent and convergent forms of thinking. Common models of design are portrayed as iterative cycles of unpacking problems, imagining and picking solutions, building and testing prototypes, and using what was learned to inform and revise the problem. Of course these models are oversimplifications because within each of the steps there are habits of mind that are necessary for engaging in important design processes that include cognitive and metacognitive tasks, creative risk-taking, failure management, and collaborative reasoning.

Good designers carry out a variety of sophisticated cognitive and metacognitive tasks. Successful designers unpack a problem and imagine differing solutions; think about potential problems and identify existing assumptions; share, build and negotiate ideas with a team of designers and with various stakeholders [27, 28]; use these insights to reflect on and decide upon the best design path; articulate their vision to others through design of artifacts [8, 28] test their existing assumptions by monitoring their designs in action; and then use insights from such evaluations to iteratively refine their designs in order to align it better with the context in which it will be used [28, 29, 24]. Much like science inquiry is dependent on metacognition in order to understand and apply underlying epistemology [32], so is design [6].

There is also an element of creative risk-taking and comfort with failure that designers must embrace to be successful. When innovators believe that failure is a bad thing they avoid taking risks, pushing their own boundaries of learning and creativity, and may inadvertently make errors that lead to more failures [22]. The creation of a work culture that promotes creative risk-taking and values the sharing of learning experience is a critical factor associated with innovation [16, 30]. Unfortunately, failure is often perceived so negatively that many avoid taking risks for fear of failing, thus preventing positive learning benefits and advancement for individuals and society [23].

Many also argue that design innovation lives in the interactions between and across people as they work to synthesize individual ideas into a collective whole and collectively negotiate what is known to create something that no one person could have accomplished alone [26, 31]. This is why collaborative teams and collaborative skills are a key part of software, engineering, and interaction design [7, 13, 27]. However, collaboration can also add many demands, both cognitive and socio-emotional [3, 17]; which may be why most people are poor collaborators, unable to engage in productive collaborative reasoning [20].

The research on design maintains that it is thoughtful, collaborative, ambiguous, and emotionally difficult process that is often at odds with traditional education. For example, effective design pedagogy promotes collaborative decision-making processes across multiple team members [13]. However, most educational environments do not help students develop these collaborative skills [5, 15, 20]. Design projects in the field are also generally long, open-ended problems [13], not short step-by-step projects that are so pervasive in traditional education. Traditional educational environments work to reduce the likelihood of failure by making it easier to learn and succeed while carrying out projects [5, 4, 2]. In addition, traditional learning environments push students to compete against each other for grades and resources and see failure as a negative outcome. These conditions interfere with the development of psychological safety, a feeling that it is safe to take interpersonal risks by suggesting new ways of thinking or admitting what is not known or that one has failed [14]. Psychological safety is recognized as a necessity for design innovation, as well as team and organizational learning [1, 14].

What the collective research on design and innovation implies is that effective design requires the practice of many complex, underlying ways of thinking that students may have little practice in carrying out. Consequently, students do not have an opportunity to develop habits of mind to successfully carryout complex design practices. So, the question arises as to how we can help students to understand and develop ways of thinking and being that align with desired design practices.

### Cognitive Apprenticeship

Many domain practices require students to carryout implicit thinking processes. Students, being novice learners may not know enough about a craft or have sufficient experience within it to carry out such processes. Moreover, they may be unaware of the habits of mind they have developed through schooling that may counter desired habits of mind for the new craft.

Cognitive apprenticeship is an approach to instruction that recognizes these problems and aims to resolve them through the application of key instructional principles [9]. These principles include specific content, methods, sequencing, and sociological considerations. Content refers to various forms of knowledge needed for the development of expertise that cover domain content as well as ways of thinking (i.e., heuristic strategies, regulation strategies, and learning strategies). Methods refer to ways of helping students develop expertise that include specific forms of instructional activity such as modeling ways of thinking and getting students to reflect on their existing habits of mind. Sequencing refers to ways of ordering activities to help students see the whole of the activity before diving into the parts, such as presenting the entirety of a design cycle and aims before unpacking each phase of design or specific techniques. The sociological component of cognitive apprenticeship refers to the type of culture that an instructor should aim to develop within the educational context. An important component of this is the creation of situated opportunities for learning: complex learning environments that require students to develop and apply knowledge in messy contexts similar to those in which they will need to apply knowledge later on.

Design is a great example of a chaotic context because it requires many complex forms of thinking. As such, it has been argued to be an optimal context for developing complex forms of thought [12, 18, 19]. Nonetheless, the most commonly cited examples of cognitive apprenticeship occur within the fields of reading, writing, and mathematics [11, 9]. Developing models of design education that build on principles of cognitive apprenticeship may help the field to ensure that the next generation of designers are well equipped to deal with the the many complex design problems that lie ahead for our society.

**EMBEDDED DESIGN PRACTICE**
We drew on principles of cognitive apprenticeship to guide the development of design curriculum and activities for an after-school club called "ThinkerSpaces Design Studio". The after-school club is intended for students in grades 4-7. It takes a playful approach towards developing design habits of mind by introducing design concepts and providing ongoing design challenges that students can solve with a variety of playful technologies. These technologies include Legos, Minecraft, Makey, Makey, littleBits, and many more.

One important aspect of our approach is our method of helping young learners develop design expertise in a situated context, which we refer to as embedded design practice. Embedded design practice is an application of the principles of cognitive apprenticeship, but tailored for design contexts. For example, when modeling, we go beyond modeling of specific techniques towards modeling of overarching practices, where students are taking part in design and development projects as both clients and designers. The ongoing narrative being that the facilitators are designers challenegd to design an after school club that develops important thinking skills in a fun and enjoyable way. This narrative allows us to model the whole design cycle: as facilitators aim to iteratively design curriculum and tools that meet students' needs, while also having opportunities to model ways of thinking and being in context at each phase in design. After such modeling, students can practice emulating these practices as part of their own projects with ongoing scaffolding, articulation, reflection, and exploration. This form of embedded practice provides opportunities for rich discussions about problems students and facilitators face with their emotions, thinking, fear of failure, collaboration, etc.

This leads to an interesting difference between our tailored approach and cognitive apprenticeship as it is commonly portrayed: the prominent role that emotion plays within the development of design expertise. As we have worked with students, we have come to understand that knowledge of emotion and the regulation of emotion can mean the difference between successful and unsuccessful design experiences. For this reason, we work to develop students' cognitive design skills while also working to develop socio-emotional skills. We help individuals and groups set emotional regulation goals, reflect on their resulting design performance and experiment with new ways to resolve emotion-related design problems. All of this is done in the service of developing an understanding of key design ideas associated with emotion, such as emotional design, empathy, perspective taking, creativity, risk-taking, and managing emotions related to failure.

Besides helping students develop a deeper understanding of design practices, embedded design practice also provides us with an opportunity to test educational designs and tools in semi-authentic settings. We say semi-authentic because testing happens in a real learning context that is outside of the lab, but we are aware that the testing is being conducted by researchers.

The testing of designs, whether it focuses on how we design the club curriculum or on the potential trade-offs of different technologies provides us with the opportunity to learn more about the needs of our students and the effectiveness of our designs while providing rich learning contexts for students to develop habits of mind. To illustrate how embedded design practice makes it possible to simultaneously learn more about the needs of students and develop their design habits of mind, we present the following example of a usability study embedded within design curriculum.

**An example of embedded design practice: Testing CoLearner:**
We wanted to use a new technology, called CoLearnr, to support and document young students' design activity. CoLearnr allows students to curate their own learning processes. This technology is akin to a learning management system (LMS). However, unlike the majority of LMSs[1], it gives instructors and students equal agency over the materials and activities housed in the system. It also allows students to click on any uploaded file, video, or website, and expand it such that a group of people can examine it and chat about it in real time in the system (see figure 1). We wanted to organize all of our instructional materials and scaffolds in to the system by phase in the design cycle and then let student teams modify and add to their own team-based instantiations. However, we did not know how easy the technology would be for students to use or whether students would know how to use all of the collaborative features. At the same time, students were beginning to show aversion to the testing phase of the design cycle.

At this point in the semester, the children had spent three months completing the first three parts of a simplified design cycle: questioning (requirements), planning (design), and building (development), while checking design quality throughout. Students were apprehensive about the fourth phase, formally testing their designs. Many students admitted they were afraid of failing, a common problem we faced with students. As such, we decided that embedding a real usability test, with a real client, would be a good way to test CoLearnr while also helping students to develop an understanding and appreciation for the purpose and techniques of design testing and the opportunities that testing to failure provide.

We began the embedded testing sessions by playing a video we created with the CEO of CoLearnr, where he explained the design problem and challenge. He explained that most LMSs used in K-12 are designed to provide the instructor with control over orchestration of the learning process and students with access to learning activities and resources. For example, administrative features (aimed at teachers) allow to control the types and quantity of resources, discourse activities, and feedback available to students. As a result, teachers are largely responsible for managing learning in the community and the resources available to it and students are largely responsible

---
[1]Google Classroom, Pearson Successnet, Haiku Learning, Agilix, Blackboard, Canvas, Schoology, Desire to learn and Moodle

iii

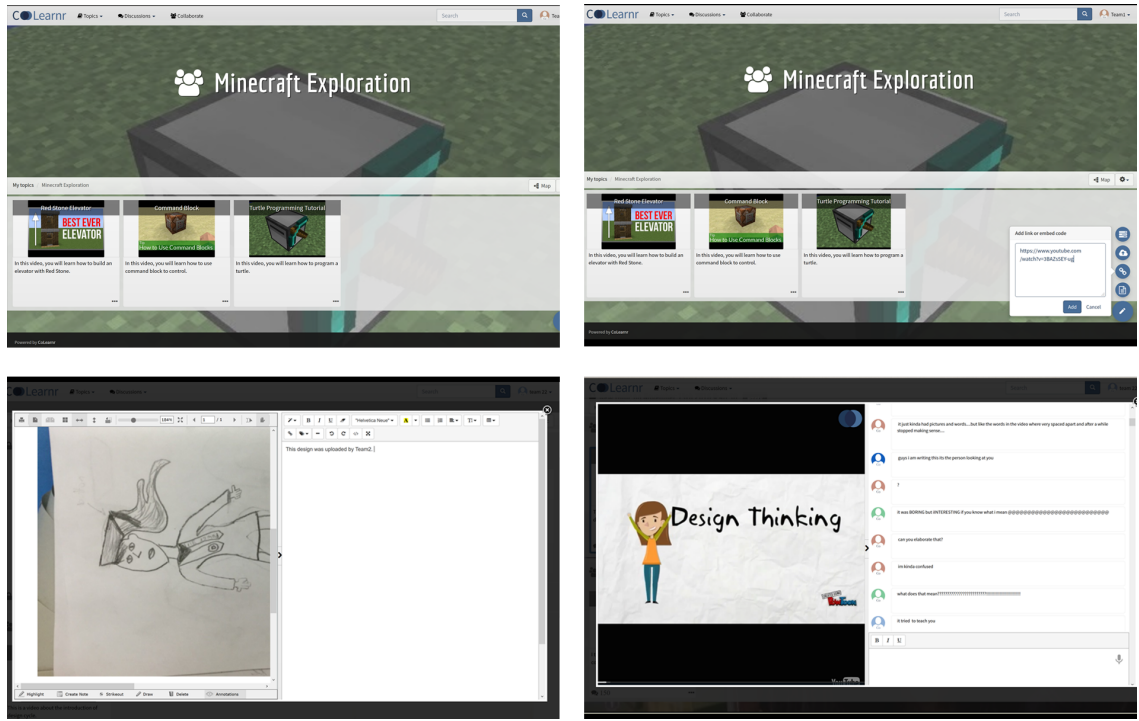

Figure 1. Clockwise from left. 1.a Home screen of CoLearnr. 1.b) Student uploading video to Minecraft topic, 1.c) Student taking notes on teammate's design, 1.d) A student's group having discussion around a video explaining design cycle

for accessing information and completing predetermined assignments. This division of labor is designed into the majority of LMSs and is problematic because such division reduces opportunities for students to practice curation skills: the ability to gather, organize, classify, and prepare digital objects for others to view and learn from. Given that the process of curation has been argued to be an increasingly important learning activity [25] and also a process of design [28], we wanted to develop a learning management tool that would allow students to take part in collaborative curation, what we refer to as a collaborative learning management systems (CLMS).

The CEO then said that he had been working on developing a collaborative learning management system. In developing this tool, he said, he followed the same careful design process that students are learning about, but nonetheless he can only assume students will know how to use his system because he has not tested it with real students yet. And so, he needed their help to test the system and help us identify all the ways it causes problems for them and fails to meet their needs so that we can make it better. He also said he was excited to find out how he could make his system better and so needed them to be completely honest. He said he knew the system was probably not that well designed yet- even though he has tried hard to make it so.

We then told the students we developed a plan to test the system, where students would carry out a series of activities, important to the system, that we assumed would be easy for them to do. If students struggled with these activities, then we would know that the design was in need of improvement ands could use information from the usability test improve it. To find out how best to meet their needs, the user needs, we said we would follow-up with a focus group. During this focus group the class could talk to the CEO in real-time so he could ask them questions to better understand what they did and didn't like and what they thought would make the system better. We then began the embedded usability test.

*Embedded Design Practice Aims*
Our main aims for this activity were to introduce students to an authentic testing experience that used real-world metrics and also get information from the usability test to improve the tool for their use. We did not expect our learners to fully understand the metrics or methods we used, but we did want them to experience the testing phase so we could refer to it later when helping them create their own plans for testing. More importantly, we wanted to model habits of mind that are important for design testing, such as testing to failure, the nature of iterative improvement, seeing failure as an opportunity to learn, looking forward to user feedback. and managing emotions related to failure.

*The Embedded Design Task*
We conducted three usability test sessions across different club lessons. Each session was used to test one of the three design features of CoLearnr. The three activities and the corresponding CoLearnr features were (1) Activity-1 Taking notes while watching a design video to test individual note-taking tools; (2) uploading a design artifact created in the previous class and sharing it with the teammates to test collaborative curating tools; and (3) Having discussion around design process

iv

video to test the collaborative chat features. During the three sessions of the usability test, we had total participation of 94% (15 students), 81% (13 students) and 88% (14 students).

The first two usability sessions lasted for 30 minutes while the third session was conducted for 50 minutes. The CEO of CoLearnr worked with us to create short introductory videos for each session. In these videos, he explained the purpose of each feature and then requested the children to provide feedback as clients. After showing the video, facilitators assigned an activity to perform using a specific CoLearnr feature. For example, uploading the pictures of the design artifacts created in the previous week (see 1.c). Each activity had a benchmark time within which the children had to complete the activity successfully. The benchmark times were set based on the prior experience and observed usage during pilot study. During the task performance, three facilitators carefully observed each group's activity.

During the activity, instructors used a pre-formatted sheet to take diary entries and collect data on specific actions of children. Apart from this, each instructor was also present to answer queries of the children while they performed the activity. We used two measures to examine usability for each of these activities *total time taken* and *accuracy*. *Total time taken* compares the time taken by the students to complete a task against a set benchmark time. *Accuracy* examines the amount of help students need to accurately complete a task. These include the number of nudges and reveals. Nudges are indirect hints provided to children upon request for assistance, "Can you see any option to upload on the page?" (facilitator's nudge when p3 was confused and requested assistance). On the other hand, reveals were the instances during the activity in which the answers were explicitly provided to the student by the instructor after the benchmark time was passed or deliberately shown by their peers. Both measures were collected from careful analysis of computer screen recordings, video of groups, and diary entries from the facilitators.

After we we finished all three usability tests, we conducted the focus-group session led by the CEO of CoLearnr. In his role as the designer, he was careful to model his eagerness to hear and understand the students' (Users') perspectives, his enthusiasm for receiving feedback, his appreciation for their time, and his reframing of design failures as opportunities to learn from the students how to make his system better. Afterwards, the facilitators talked about how it must feel to work hard on a design only to find flaws, but how important it is to take the perspective that Prabhu did, to be thankful for the opportunity to learn. This experience helped us to discuss productive failure and prepare students for the following weeks, when they would test their own designs.

## OPPORTUNITIES FOR LEARNING PROVIDED BY EDP

### Understanding students' technology needs

The facilitators and the primary designer of the Colearnr system gathered information about the students from the embedded study. The embedded study showed us that our seemingly technology savvy students were actually quite naive when it came to certain technology related tasks. Of the three tasks they were required to carry out in CoLearner, they experienced most difficulty with uploading content. This task took students the longest to complete and required the most nudges (scaffolding) out of the three tasks (see Tables 1 and 2). Only four students (31% of our students) were able to complete within the benchmark time and five students were unable to complete the activity.

| Activities | Benchmark time (min) | Time taken (min) Mean | S.D. |
|---|---|---|---|
| Note taking | 4.00 | 4.23 | 3.43 |
| Uploading | 5.50 | 5.35 | 3.23 |
| Discussion | 3.00 | 1.85 | 1.50 |

**Table 1. Performance values for all the activities**

Our qualitative data gave few insights to explain students difficulties. The majority of participants revealed that they found the task of uploading activity confusing. For instance, one child was observed replying to a facilitator on providing a nudge - *"I don't know what a Desktop is..."*. Understandably, she required more nudges than the mean value. Despite receiving constant nudging, considerable number of children made mistakes while uploading. The most common was trying to upload the image files by clicking on the share button in the image application (Preview on Macintosh). Other mistakes included copy and pasting or dragging and dropping the image in the wrong place (such as in the discussion box).

Students also experienced some difficulty with the note-taking activity. 60% of the participants completed within the set benchmark whereas 73% of the participants completed the task irrespective of the benchmark (figure 2).

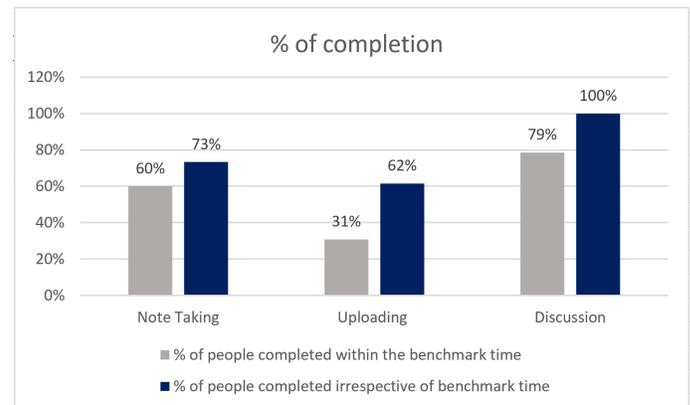

**Figure 2. Performance data depicting % of people who completed the task within and irrespective of benchmark time.**

| Activities | Number of Nudges Mean | S.D. |
|---|---|---|
| Note taking | 1.50 | 1.02 |
| Uploading | 2.85 | 1.63 |
| Discussion | 0.14 | 0.36 |

**Table 2. Performance values for all the activities**

Four students were not able to complete the given activity. The children who were able to complete the activity took mean time of almost four and half minutes (n=15) (table1). This was a little more than the benchmark time which we had provided for the children. Among the participants, the maximum time taken to complete the first activity successfully was six minutes. Majority of the students who completed the activity used the note-taking function to take notes while seeing the design video simultaneously. However, 47% of the children revealed that the note taking activity was confusing.

> I think it was kind of confusing entire way through (referring to notes), figuring out where it was and how to get to it. Did not get much info on how to use it. - P12

The confusion also extended to the understanding whether they felt the notes were private. A little less than half of these students expressed confusion when we asked whether they thought the notes they took were private. Despite this, almost half the students (45%) liked the process of taking notes. On being asked why P02 mentioned that *"It is easier to type than to write it out on a paper"*. All of these students were using advanced features in the notes which included changing font style and colors, using copy and paste, and editing the HTML code of the WYSIWYG (What you see is what you get editor) editor.

> [...] I copied and pasted a lot. Sometime, I messed up bit what I already did so I copied it.- P08

The easiest activity for them was participating in an online chat discussion while watching a video, where 79% of the children completed the task within the benchmark time and 100% completed the activity. The high numbers in the performance data were also reflected in the focus group. All the students found the discussion feature to be easy and intuitive. Many of them (57%)even expressed explicit verbal likeness towards the feature.

> [...] I liked discussion better because it allows you to talk with friend, even if it is not your group. I could talk to the groups which are across the room. - P07

In making sense of our findings, we were aware of the many restrictions that these students had when using technology in the school. Technological experiences were restricted due to the limited resources, strictly defined curriculum, and lack of guidance. A primary concern with their computer-based experiences was preventing issues related to privacy, computer viruses, and access to inappropriate content. Thus, it was not surprising that our students had more knowledge related to discussing and downloading information from predetermined websites than they had about creating information and sharing artifacts. What was surprising, was student's lacking capabilities to handle computer activities outside the browser. Students were not aware of terms such as desktop, image viewer, and finder. Children were also unfamiliar with performing certain computer operations, including drag and drop, increasing the thumbnail size in the finder, locating the desktop folder in the finder and selecting images in the image viewer.

These findings challenge the notion that students as such are tech savvy as many presume them to be. It is also problematic because it suggests that the primary ways in which our learners were using computers was for lower forms of cognition. Though we were encouraged to continue developing CoLearnr for use with our students, we also learned that we needed to develop the tool with consideration to the lack of skills students have when it comes to curation.

**Situated Learning Experiences**
While we were learning more about the technological experience and needs of our learners, they were able to see what testing looked like and shift perspectives from a designer to a user. We were also able to model important designerly ways of being that we could refer to later on.

From a designer's perspective, students discussed being hesitant to get feedback from others for fear of failing or losing face, i.e., seeming less than in the eyes of their peers. This is understandable because many students do not know how to frame feedback in a constructive way. However, in working with the CEO of CoLearnr, students were able to experience testing, watch him model feedback seeking, work on giving constructive criticism, and discuss with the facilitators how they felt about the experience and how the CEO must have felt.

Being in the role of the provider of feedback allowed students to feel valuable and understand how important the role of the user is to a designer. The students discussed how enjoyable the experience was and how they tried to frame their feedback in ways that would not hurt teh designer's feelings but also help him understand the utility of his design. This helped us to realize how difficult it was for students to give and receive feedback, and the need for us to develop games and scaffolds to help students develop these skills.

These discussions and experiences were also important cognitive tools for future skill development and reflection. They were common reference points that students would bring up in later sessions as they prepared to get and give feedback. During reflections that occurred during the testing phase. We discussed how interesting it can be to get feedback, to become aware of problems you did not know existed, and how identifying these problems makes you a better designer. We discussed how good designers, like the CEO of ColLearnr, keep testing their designs until they fail because, only then can they identify potential problems and fix them. Facilitators would also draw on these experiences as a mean to contextualize important skills and discuss their value to design. For example, we drew on this embedded experience to discuss why empathy is important, why testing our designs with real people is important, and why failing is a great thing in design and engineering- because it gives us opportunities to learn.

**CONCLUSION**
As learning technologies are becoming more integral to education, it is important that we create more technologically meaningful designs. The current study was an attempt towards bringing together the technology designer and teacher to create those meaningful learning environments in collaboration with each other [10]. This embedded nature of design, where design is both a means and an end in itself, empowers teachers to be an equal participant in designing meaningful products. In addition, such settings enable teachers to integrate various learning methods contextually within various design activities to improve situated learning experiences for children.

The design conversations between the CEO (the designer) and the students (the users), within a design learning context, provides authentic conversations that play a dual role. One, they act as fodder for design improvements for the designer and second, they become learning objects for making sense of design processes during whole class reflections or ongoing coaching. As these conversations become objects to reflect and think about, they make the cognitive processes behind designing more accessible to the learner, breaking down the complexity of the design process. Approaches like embedded design practice in essence are trying to bring the integrative thinking skills in the forefront, where the designer is required to make a series of decisions, through cognitive and metacognitive processes to solve authentic problems.

Embedded design practice can also act as a medium for designers to take part in enculturating young designers to the craft of design, while learning more about their needs. This meaningful context, allows designers to immerse themselves in a two way dialogue to model design practice and become more aware of young learners' mental model and design gaps which might have been easily overlooked.

An important future line of research for design education in K-12 settinhgs, is to examinine the differences between the teacher moves used by expert and novice facilitators in these complex design contexts in order to identify the professional development needs of novice design teachers. Through this work, we can begin to develop more robust teacher development programs that help to make design habits of mind and the techniques used develop them more accessible to teachers. Such lines of research could help to ensure that K-12 education succeeds in creating authentic design experiences to support the next generation of human-centered designers.